\documentclass[12pt]{article}
\usepackage{amssymb}

\usepackage{graphicx}
\usepackage{epstopdf}
\usepackage{sw20aip}

\input{tcilatex}
\begin{document}

\author{Manfred Bucher \and Physics Department, California State University, Fresno
\and Fresno, CA 93740-8031}
\title{Coulomb oscillations as a remedy for the helium atom }
\date{October 8, 2007 }
\maketitle

\begin{abstract}
The largest failure of the old, Bohr-Sommerfeld quantum theory was with the
helium atom. It brought about the theory's demise. I show that this failure
does not originate, as commonly believed, with the orbit concept \textit{per
se.} Instead, it was caused by the wrong choice of orbits, compounded by
ignorance of the exclusion principle. Choosing semiclassical electron
oscillations through the $He$ nucleus, I calculate a singlet ground-state
energy that rivals in accuracy with quantum-mechanical results. The same
method reveals Bohr's historic energy value as the forbidden triplet ground
state---a result beyond the reach of quantum mechanics. At the qualitative
level, the concept of Coulomb oscillations visually explains the major
features in the $He$ double spectrum in terms of crossed or parallel orbit
orientation.\bigskip

\noindent \textbf{Keywords:} Atomic electron structure, Semiclassical
theories, Helium spectrum, History of quantum theory
\end{abstract}

\section{INTRODUCTION}

The helium atom broke the back of the old quantum theory. The approach that
had started so promising with the Bohr model of 1913 was finally abandoned a
decade later. Already in his seminal trilogy, Bohr extended his model of the
hydrogen atom to $He$ (Bohr 1913) such that two electrons would orbit---at
diametrically opposite positions---a nuclear double charge on one circle
(see Fig. 1). The calculated ground-state energy was of the right order of
magnitude---5\% off the measured value. Although an encouraging result, it
was a far cry from the ``spectroscopic'' accuracy of the energy terms that
the Bohr model gave for the (one-electron) hydrogen atom. Therefore efforts
were soon made by a host of researchers---Sommerfeld, Land\'{e}, Kramers and
Bohr, Van Vleck, Pauli and Born, Heisenberg and Born (Mehra and Rechenberg
1982a)---with improved models of the $He$ atom. Among the numerous attempts
were a distinction of separate (inner and outer) coplanar electron orbits,
then elliptical orbits in tilted planes subject to various phase relations.
Later, perturbation theory from celestial mechanics was employed. However,
despite increasing efforts and sophistication, those calculations did not
converge toward the experimental value. In fact, the most extensive methods,
carried out by Kramers and by Van Vleck (Mehra and Rechenberg 1982a), gave
results that disagreed about as much (+5\%) from the experimental
ground-state energy as Bohr's original, ``two-seat-roundabout'' model
(-5\%). Much worse results were obtained by Born and Heisenberg for the
excited states of $He$. In despair, the old quantum theory of orbit
quantization was forgone in the mid 1920s and a search for new principles
ensued. This led to Heisenberg's discovery of matrix mechanics. Among the
early triumphs of the new quantum mechanics were easy calculations of the
ground-state energy of $He$ with perturbation theory (relative deviation $%
\Delta $ $\approx $ 5\%) and variational techniques ($\Delta $ \TEXTsymbol{<}
2\%), and Heisenberg's explanation of the origin of the $He$ singlet and
triplet spectra (Mehra and Rechenberg 1982b).\bigskip
\begin{equation}
\includegraphics[width=2in]{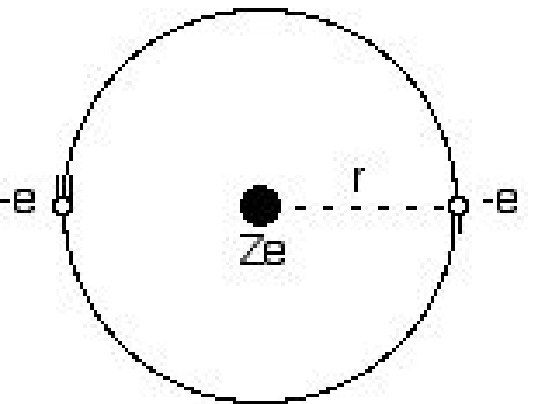}   \tag{}
\end{equation}
\begin{quote}
Fig. 1.  Bohr's historical 
"two-seat-roundabout" 
model of the  $He$ atom. \bigskip
\end{quote}

The helium atom was not the only failure of the old quantum theory. Two
other short-comings were the theory's failure to give the correct multiplet
structure of the hydrogen atom (neglecting spin) and the stability of the
hydrogen molecule ion, $H_{2}^{+}$. However, as I have recently shown
(Bucher 2005), both these dilemmas can be resolved, within the framework of
semiclassical orbit quantization, by including oscillations of the electron
through the atomic nucleus (molecular nuclei). I call them \textit{Coulomb
oscillations}. After those successes it is reasonable to extend the
Coulomb-oscillator concept to the $He$ atom---a two-electron system. It is
then found that the calculated energy for the singlet ground state is
comparable with that from quantum mechanics. The same method yields the
energy of the forbidden triplet ground state---a result outside the scope of
quantum mechanics. The Coulomb oscillations also provide, through orbit
visualization, a simple explanation of the symmetry splitting and the energy
distinctions of the excited states in the $He$ spectrum.

\section{BOHR'S HELIUM MODEL}

In Bohr's historic model of the $He$ atom, the centripetal force on each
electron is 
\begin{equation}
\frac{mv^{2}}{r}=\frac{Ze^{2}}{r^{2}}-\frac{e^{2}}{(2r)^{2}}\equiv \frac{%
Z^{\prime }e^{2}}{r^{2}}.
\end{equation}
The electron-electron repulsion can be formally combined with the nuclear
charge (number) Z such as if each electron were subject to only a central
force from an \textit{effective} nuclear charge, 
\begin{equation}
Z^{\prime }=Z-\frac{1}{4}.
\end{equation}
The orbit is then quantized by the requirement that each electron possesses
an angular momentum 
\begin{equation}
mvr=n\hslash ,
\end{equation}
where $\hslash =$ $h/2\pi $ is the reduced Planck constant and $n$ a quantum
number. Solving Eqs. (1) and (3) for $v^{2}$, and equating, yields the
quantized orbit radius, 
\begin{equation}
r_{n}=\frac{n^{2}}{Z^{\prime }}r_{B}.
\end{equation}
Here the Bohr radius, $r_{B}=\hslash ^{2}/me^{2}=0.53\times 10^{-10}m$,
serves as a universal atomic length unit. The orbit energy of \textit{both}
electrons, 
\begin{equation}
E=2\times \frac{1}{2}mv^{2}-\frac{2Ze^{2}}{r}+\frac{e^{2}}{2r}=mv^{2}-\frac{%
2Z^{\prime }e^{2}}{r},
\end{equation}
becomes quantized, after insertion of $v^{2}$ and $r_{n}$ from Eqs. (3) and
(4), 
\begin{equation}
E_{n}=-\frac{2Z^{\prime 2}}{n^{2}}R_{y},
\end{equation}
expressed, for convenience, in terms of the Rydberg energy unit, $R_{y}$ $%
=me^{4}/2\hslash ^{2}=13.6$ $eV$.

\medskip From the formalism above one can see that Bohr's
``two-seat-roundabout'' model of the $He$ atom is merely a $Z^{\prime }$
-scaled version, Eq. (2), of his model of the $H$ atom. The ground-state
energy of the $He$ atom is then 
\begin{equation}
E_{1}=-2Z^{\prime 2}R_{y}.
\end{equation}

\begin{equation}
\includegraphics[width=3in]{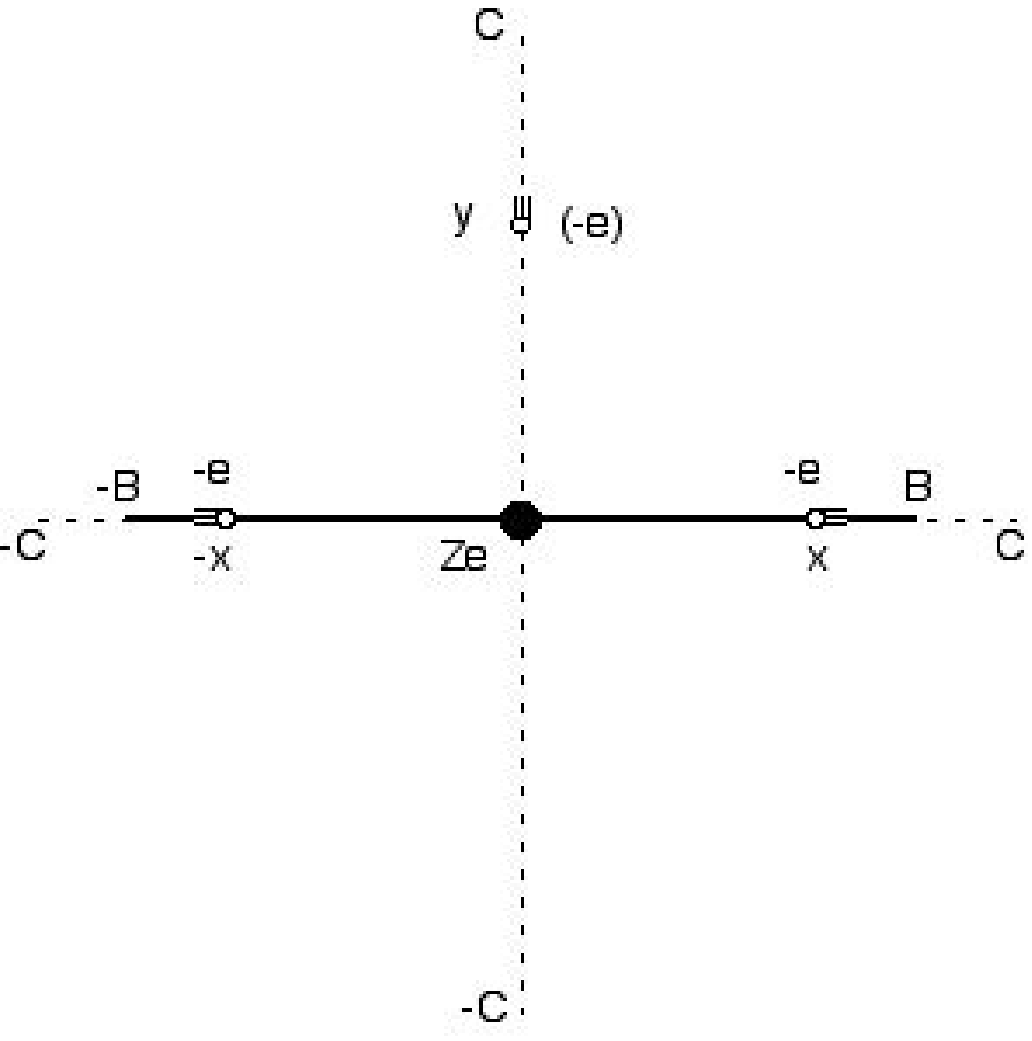}   \tag{}
\end{equation}
\begin{quote}
Fig. 2. Contralinear (bold) and cross-synchronous (dashed) Coulomb oscillations of two electrons through a nucleus.\bigskip
\end{quote}

\section{COUPLED COULOMB OSCILLATORS}

First assume that both electrons swing in opposite phase along the same axis
through the nucleus, always at mirror positions, $x_{1}=-x$ and $x_{2}=x,$
with respect to the nucleus at $x=0$ (see Fig. 2). I call this oscillation
mode \textit{``contralinear.''} The atom's total energy $E_{-xx}$ at any
mirror position $x$ of the electrons must equal the potential energy of both
electrons at their turning points, $-B$ and $B$, 
\begin{equation}
E_{-xx}=2\times \frac{1}{2}mu^{2}-\frac{2Ze^{2}}{|x|}+\frac{e^{2}}{|2x|}=-%
\frac{2Ze^{2}}{B}+\frac{e^{2}}{2B}\equiv -\frac{2Z_{-xx}^{\prime }e^{2}}{B}
\end{equation}
with 
\begin{equation}
Z_{-xx}^{\prime }=Z^{\prime }=Z-0.25
\end{equation}
from Eq. (2). Here $u$ denotes each electron's speed in the contralinear
oscillation, 
\begin{equation}
u(x)=\sqrt{\frac{e^{2}}{m}}\sqrt{\frac{2Z}{|x|}-\frac{1}{|2x|}-\frac{2Z}{B}+%
\frac{1}{2B}},
\end{equation}
and, after simplification, 
\begin{equation}
u(x)=\sqrt{\frac{2Z_{-xx}^{\prime }e^{2}}{m}}\sqrt{\frac{1}{|x|}-\frac{1}{B}}%
.
\end{equation}

The action integral over an oscillation cycle must, by Sommerfeld's
quantization condition, be an integer multiple of Planck's constant, 
\begin{equation}
A_{-xx}=m\oint u(x)dx=4m\int_{0}^{B}u(x)dx=nh,
\end{equation}
with $n=1$ for \textit{each} electron in the $He$ ground state. The integral
in Eq. (12) can be solved analytically (Bucher 2005). This then gives the
oscillation amplitude, 
\begin{equation}
B_{n}=\frac{2n^{2}}{Z_{-xx}^{\prime }}r_{B},
\end{equation}
and, after insertion into Eq. (8), the same ground-state energy as in Eq.
(7), 
\begin{equation}
E_{-xx}=-2Z_{-xx}^{\prime 2}R_{y}.
\end{equation}
The contralinear Coulomb oscillator is thus energetically equivalent to
Bohr's two-seat-roundabout model of the $He$ atom: In both cases the
electron-electron distance $\overline{-xx}$ is twice the (varying or,
respectively, constant) electron-nucleus distance, 
\begin{equation}
\overline{-xx}=2x.
\end{equation}

Let us next explore synchronous Coulomb oscillations of two electrons
through the $He$ nucleus in \textit{perpendicular} directions and with equal
amplitude $C$ (dashed in Fig. 2). Again, the atom's total energy $E_{xy}$
for electron positions mirrored off the $y=x$ diagonal, $\mathbf{r}%
_{1}=(x,0) $ and $\mathbf{r}_{2}=(0,y)$, must equal the potential energy of
both electrons at their turning points, $x=C$ and $y=C$, 
\begin{equation}
E_{xy}=2\times \frac{1}{2}mw^{2}-\frac{2Ze^{2}}{|y|}+\frac{e^{2}}{|y|\sqrt{2}%
}=-\frac{2Ze^{2}}{C}+\frac{e^{2}}{C\sqrt{2}}\equiv -\frac{2Z_{xy}^{\prime
}e^{2}}{C},
\end{equation}
with a new effective nuclear charge number, 
\begin{equation}
Z_{xy}^{\prime }=Z-\frac{\sqrt{2}}{4}\approx Z-0.35.
\end{equation}
Its lesser amount, $Z_{xy}^{\prime }$ $<Z_{-xx}^{\prime }$, is a consequence
of the closer electron-electron distance, 
\begin{equation}
\overline{xy}=x\sqrt{2}.
\end{equation}

\noindent compared to Eq. (15). Since Eq. (16) scales, via the effective
charge, with Eq. (8), the ground-state energy of this \textit{%
crossed-synchronous} Coulomb oscillation is analogous to Eq. (7), 
\begin{equation}
E_{xy}=-2Z_{xy}^{\prime 2}R_{y}.
\end{equation}

\begin{equation}
\includegraphics[width=3in]{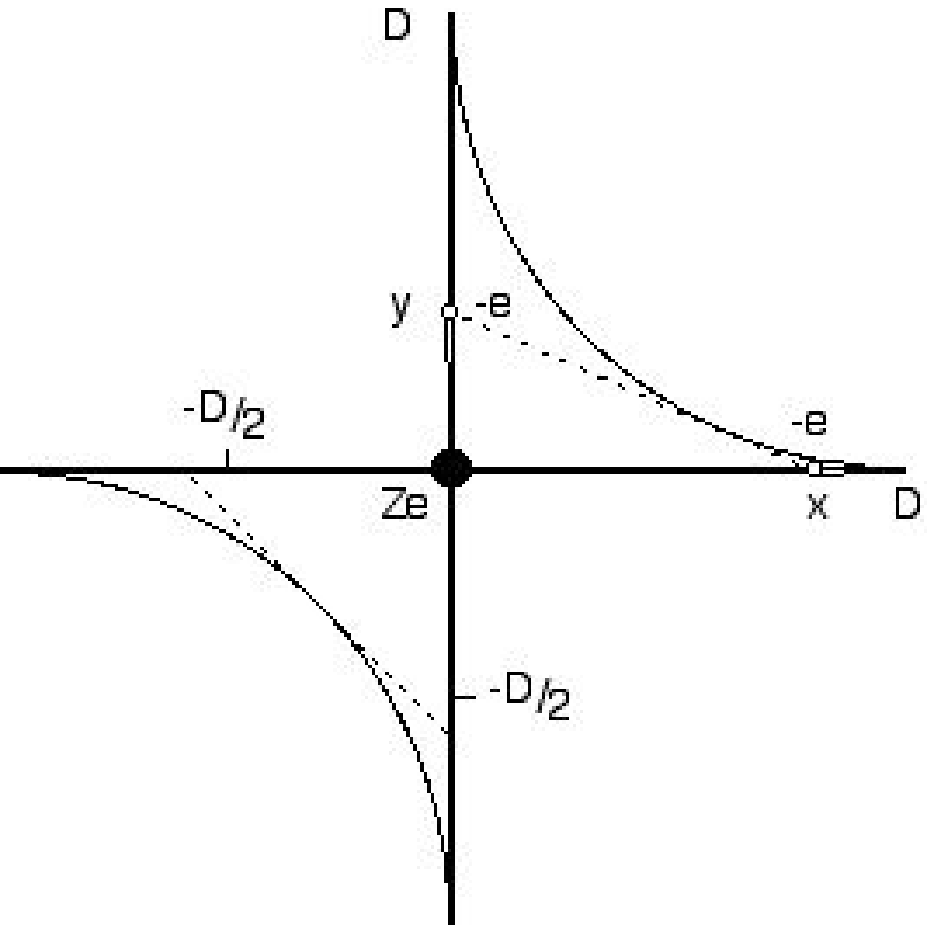}   \tag{}
\end{equation}
\begin{quote}
Fig. 3. Cross-delayed Coulomb oscillations of two electrons through a nucleus.  The (dashed) distance-line between the electrons is tangent to an asteroid curve.  The third quandrant serves in assessing the average electron-electron distance.\bigskip
\end{quote}

Lastly, consider crossed Coulomb oscillations that are out of phase by a 
\textit{quarter period} (see Fig. 3). When now one electron is at a turning
point (say, $x=D$), then the other electron traverses the nucleus ($y=0$) in
the perpendicular ($y$) direction. I denote this \textit{``cross-delayed''}
oscillation mode by the suffix $x/y$. The previous method of equating the
total energy at any position with the potential energy of \textit{both}
electrons at the turning points, Eqs. (8) and (16), no longer suffices.
However, averaged over a period, we can still employ the two-body notion of
an effective nuclear charge, $Z_{x/y}^{\prime }$, acting exclusively on each
electron as if it comprised both the true nuclear attraction and the
repulsion from the other electron. With respect to the energy, the
time-average distance from the fixed nucleus to an electron in Coulomb
oscillation is \textit{half} the latter's amplitude (Bucher, Elm and Siemens
1998; Pauling and Wilson 1935). For a pair of contralinear oscillators, the
average electron-electron distance is, by Eq. (15), twice that value, $%
\left\langle \overline{-xx}\right\rangle =2$ $\times $ $B/2=B$, but for
cross-synchronous oscillators less than the amplitude , $\left\langle 
\overline{xy}\right\rangle =\surd 2$ $\times $ $C/2=C/\surd 2$, Eq. (18).
For cross-\textit{delayed} oscillators, the distance-line between the
electrons (dashed in Fig. 3) is tangent to an asteroid curve. The \textit{\
average} electron-electron distance of cross-delayed oscillators, $%
\left\langle \overline{x/y}\right\rangle $, must be shorter than the
amplitude, $D$, but longer than the closest separation, indicated in the
third quadrant of Fig. 3. It seems safe to assume that $D/\surd
2<\left\langle \overline{x/y}\right\rangle <D$. Therefore the average of the
bracketing effective charges may serve as a reasonable approximation for the
effective nuclear charge in cross-delayed Coulomb oscillations, 
\begin{equation}
Z_{x/y}^{\prime }\approx \frac{Z_{-xx}^{\prime }+Z_{xy}^{\prime }}{2}=Z-%
\frac{1+\sqrt{2}}{8}\approx Z-0.30,
\end{equation}

\noindent and accordingly for the energy, 
\begin{equation}
E_{x/y}\approx -2Z_{x/y}^{\prime 2}R_{y}.
\end{equation}

\section{GROUND-STATE RESULTS}

Table I shows for several nuclear charges $Z$---and Fig. 4 for the special
case of $He$---experimental values and calculations of the ground-state
energy. [A simple formula for the ground-state ionization potential of $He$
and two-electron ions has recently appeared in these pages (Elo 2007).] The
calculated energies $E_{xy}$ of the cross-synchronous oscillator deviate
from experiment about as much as the $E_{-xx}$ values of the contralinear
oscillator, $|E_{xy}-E_{\text{\textit{\ expt}}}|/|$ $E_{\text{\textit{\ expt}%
}}|$ $=$ $\Delta _{xy}$ $\approx \Delta _{-xx}=$ $|E_{-xx}$ $-E_{\text{%
\textit{expt}}}|/$ \TEXTsymbol{\vert}$E_{\text{\textit{expt}}}|$, although
in opposite directions, $E_{-xx}$ $<E_{\text{\textit{expt}}}$ $<E_{xy}$.
Interestingly, these values are similar (or equal) to historical results
from the old quantum theory: The energy $E_{xy}(He)=-5.42$ $R_{y}$ is close
to the values obtained with enormous efforts by Kramers ($-5.52$ $R_{y}$)
and by Van Vleck ($-5.53$ $R_{y}$) (Mehra and Rechenberg 1982a) and $E_{-xx}$
agrees, as noted, with $E_{1}$ of Bohr's two-seat-roundabout model.

\begin{equation}
\includegraphics[width=4in]{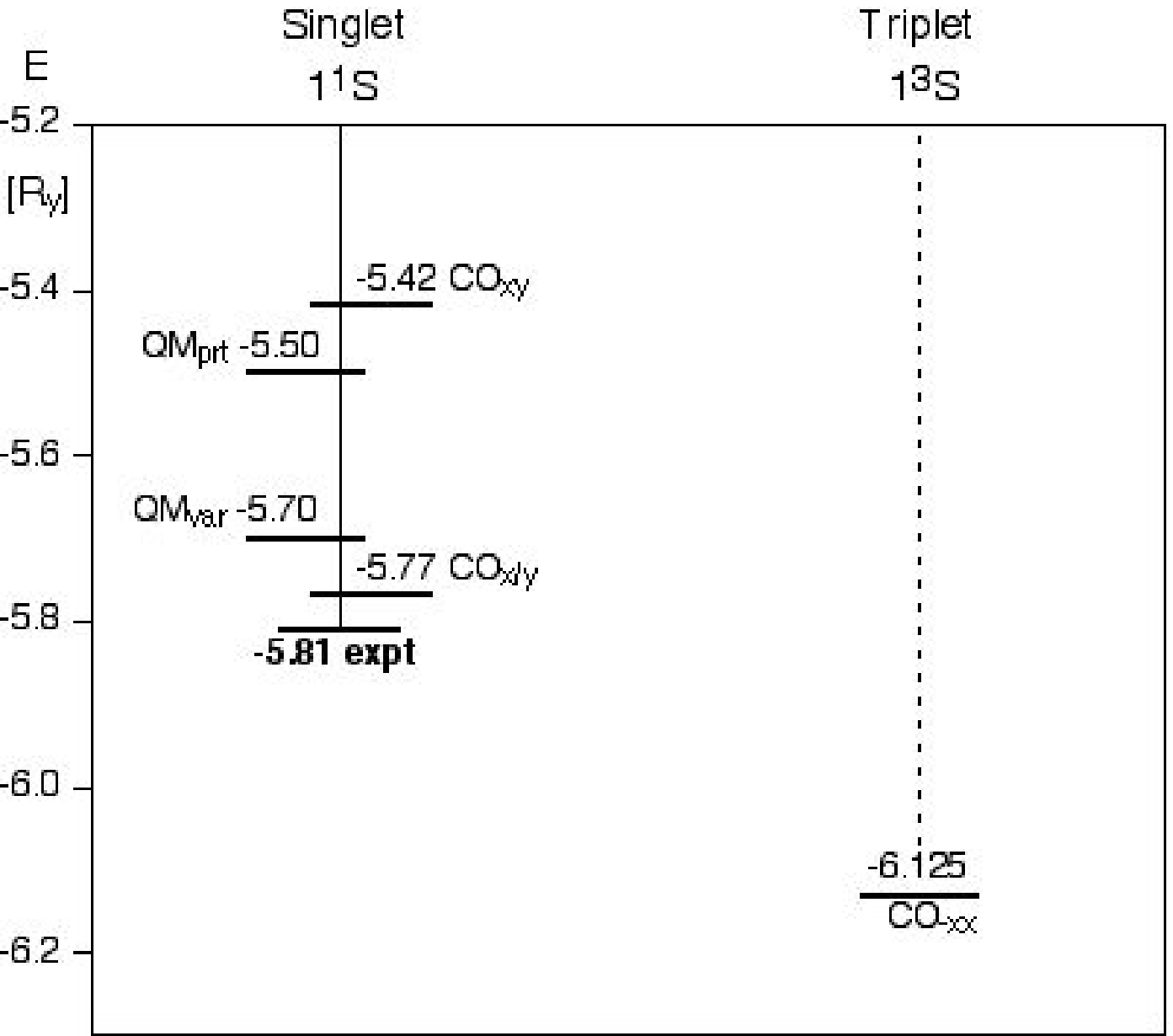}   \tag{}
\end{equation}
\begin{quote}
Fig. 4.  Experimental energy level (in  $R_{y}$) of the helium ground state and calculated values with perturbational and variational quantum mechanics (QM) and with Coulomb oscillations (CO) in contralinear (-xx), cross-synchronous (xy), and cross-delayed (x/y) modes.\bigskip
\end{quote}

The energy of the \textit{cross-delayed} Coulomb oscillations is found to be
remarkably close to the experimental data, $E_{x/y}\cong E_{\text{\textit{\
expt}}}$. Table I shows deviations of $0.5\%$ for $He$, $0.2\%$ for higher-$%
Z $ ions, and about $6\%$ for the hydride ion, $H^{-}$. The reason for the
large deviation of $H^{-}$ is the relatively stronger electron-electron
interaction which gives rise to a radius-dependent screening of the nuclear
charge, $Z-s(r),$ instead of a constant screening factor $s$ as in all the
other cases (Bethe and Salpeter 1957).

Two instructive comparisons are provided by \textit{quantum-mechanical}
expressions. First-order perturbation theory (Pauling and Wilson 1935) gives 
\begin{equation}
E_{QM}^{prt}=-(2Z^{2}-\frac{5}{4}Z)R_{y}.
\end{equation}
The perturbation values $E_{QM}^{prt}$ are about $5\%$ less binding (higher
level) than the experimental values $E_{\text{\textit{expt}}}$ but slightly
more binding than those of the cross-synchronous Coulomb oscillator, $E_{xy}$
. A one-parameter variational treatment (Pauling and Wilson 1935) gives 
\begin{equation}
E_{QM}^{var}=-2(Z_{QM}^{\prime var})^{2}R_{y},
\end{equation}
with 
\begin{equation}
Z_{QM}^{\prime var}=Z-\frac{5}{16}\approx Z-0.31.
\end{equation}
Table I shows that the variational values $E_{QM}^{var}$ deviate slightly
more from experiment than those of the \textit{cross-delayed} Coulomb
oscillator, $E_{x/y}$.

\section{INTERPRETATION}

To understand the results, a reflection on the abilities and deficiencies of
the Bohr-Sommerfeld quantum theory is in order. The triple success of the
orbit-based, old quantum theory is its correct description of the $H$ atom
(and one-electron ions) concerning (1) the energy levels $E_{n}$, (2) the
orbital angular momenta $\mathbf{L}_{nl}$---if corrected as $%
L_{nl}^{2}=l(l+1)\hbar ^{2}$ and with the Coulomb oscillator included---and
(3) the orbits' space quantization---with $(\mathbf{L}_{nl})_{z}$ $=$ $m_{l}$
$\hbar $. These three achievements are succinctly represented by the
corresponding (principal, angular and magnetic) quantum numbers: $n,l$, and $%
m_{l}$.

The shortcomings of the old quantum theory lie in its neglect of three
particle properties: their spin, their wave nature and their quantum
statistics. Inclusion of spin---expressed by the spin quantum number $s$%
---is necessary to account for the \textit{total} (instead of only orbital)
angular momentum. The wave nature---expressed in Schr\"{o}dinger's
formulation of quantum mechanics by the wavefunction $\psi $---is necessary
to account for the particles' spatial (probability) distribution, $\psi \psi
^{*}=|\psi |^{2}$. Quantum statistics becomes relevant for systems with
identical (indistinguishable) particles---like the \textit{two} electrons
about the nucleus in the $He$ atom.

In the framework of quantum mechanics, including Fermi-Dirac quantum
statistics and the related Pauli exclusion principle, the $He$ atom is
described in terms of both spatial wavefunctions $\psi (\mathbf{r}_{i},%
\mathbf{r}_{j})$ and spin wavefunctions $\alpha (i)\beta (j)$ which depend
on the electrons' position and spin orientation, respectively, and whose
product is antisymmetric with respect to an interchange of the electrons, $%
i\leftrightarrow j$ (Bethe and Salpeter 1957). More specifically, an
interchange of electrons in corresponding linear combinations of such
wavefunctions must leave the spatial part the same, but change the sign of
the spin part, and vice versa. In the first case, the combination of spatial
wave functions is said to be \textit{symmetric} and the spin-wave
combination \textit{antisymmetric}. The opposite holds in the second case.
In accordance with the quantum-mechanical multiplicity of angular momentum,
the combination of spatial wavefunctions, 
\begin{equation}
\psi _{ab}^{\pm }(\mathbf{r}_{1},\mathbf{r}_{2})=\frac{1}{\sqrt{2}}\left[
\phi _{a}(\mathbf{r}_{1})\phi _{b}(\mathbf{r}_{2})\pm \phi _{a}(\mathbf{r}%
_{2})\phi _{b}(\mathbf{r}_{1})\right] \text{,}
\end{equation}

\noindent are then symmetric (+) if the electrons' spin is opposite (singlet
state) and antisymmetric (-) when their spins are parallel (triplet state).
(Disregarding spin-orbit coupling, the terminology ``singlet'' and
``triplet'' refers to the non-splitting and threefold splitting,
respectively, of such levels in a magnetic field.) Here $\phi _{a}$ and $%
\phi _{b}$ are one-electron wavefunctions. Optical transitions occur only
between states of the same symmetry. Hence the observed $He$ double
spectrum, arising from two sets of either singlet or triplet energy levels,
as was first explained by Heisenberg.

The probability density distribution $|\psi _{ab}^{\pm }|^{2}$ of the $He$
atom contains, by Eq. (25), both atomic orbital density terms $|\phi
_{a}|^{2}$ and $|\phi _{b}|^{2}$, and so-called ''interference terms,'' $%
\phi _{a}\phi _{b}$. The latter give rise to increased or decreased electron
density in the overlap region of the $\phi _{a}$ and $\phi _{b}$ orbitals
according to symmetry $(+)$ and antisymmetry $(-)$ of $\psi _{ab}^{\pm }$,
respectively. In this way the $\pm $ symmetry distinction corresponds to
spatially contracted or expanded states of the $He$ atom.

Since (Fermi-Dirac) quantum statistics requires that all spatial
wavefunctions of the triplet state be antisymmetric, this rules out, by Eq.
(25), the $1^{3}$S ground state, $\phi _{100,100}^{-}$ $=0$. With that
symmetry property inherent in the formalism, quantum mechanics cannot give
the energy of the (not naturally occuring) $1^{3}$S state of $He$. The old
quantum theory, in contrast, does not account for quantum statistics at all.
It does therefore yield symmetry-forbidden quantum states which have to be
ruled out ``by hand.''

The spatially anti-symmetric $1^{3}$S state is semiclassically represented
by contralinear Coulomb oscillations $(-xx)$---an association borne out by a
unifying pattern that comprises both the ground state and the excited states
in the $He$ spectrum, shown below. The energy of the $1^{3}$S state of $He$
is thereby $E_{-xx}$ $=-6.125R_{y}$---Bohr's historic value for the
two-seat-roundabout model of $He$!

How \textit{realistic} is the attribution of the energy level $E_{-xx}$ to
the forbidden triplet ground state? I offer two plausibility arguments: 
\textit{Qualitatively}, it would extend the pattern of a higher singlet than
triplet level from all the excited states to the observed and forbidden
ground state, $E_{\exp t}$ $>$ $E_{-xx}$ (see Figs. 4 and 5). How accurate,
then, is its \textit{quantitative} value, $E_{-xx}$ $=-6.125R_{y}$?

Recall that both the Bohr model and the simple Coulomb oscillator (Bucher
2005) give, like quantum mechanics, the correct (non-relativistic) \textit{\
energy} of the hydrogen atom. The reason why these energy calculations are
successful is that the representation of such periodic processes in phase
space encloses the same volume \textit{(nh)}, despite the different (and
unrealistic) geometries of the semiclassical orbits compared to the
quantum-mechanical orbitals. Likewise it can be expected of the extensions
of both the Bohr model and the Coulomb oscillator to the $He$ atom, that
they give (at least) the correct ground-state energy: For once, Bohr's
two-seat-roundabout model, and equivalently the contralinear Coulomb
oscillations $(-xx)$, are reduced, via scaling, to the (correctly solvable)
one-electron problem. Furthermore, they represent the largest (average)
electron-electron separation which yields the lowest energy (ground state).
Nevertheless, the semiclassical $1^{3}$S state must be discarded because of
its forbidden quantum symmetry, or equivalently, its violation of the
exclusion principle---a constraint beyond the old quantum theory.

The \textit{experimentally} observed ground-state of the $He$ atom is the
remaining $1^{1}$S singlet state. By Eq. (25), its wavefunction $\phi
_{100,100}^{+}$ is spatially symmetric. Semiclassically, it is represented
by \textit{crossed} Coulomb oscillations which possess higher (fourfold)
symmetry than the (twofold) mirror symmetry of the linear oscillations. When
the cross-oscillations are synchronous $(xy)$, then the lesser electron
separation, Eq. (18), causes too high an energy level, $E_{xy}>E_{\exp t}$.
The larger \textit{average} electron separation of the cross-\textit{delayed}
oscillations, indicated in Fig. 3, lowers the energy level accordingly, $%
E_{x/y}\cong E_{\exp t}$. Incidentally, those two types of \textit{cross}
-oscillations can be regarded as the semiclassical equivalent of the
quantum-mechanical perturbative and, respectively, variational treatment
which similarly differ in the average electron separation (Bethe and
Salpeter 1957). Such correspondence is reflected in comparable energy
levels, $E_{xy}\approx E_{QM}^{prt}$ and $E_{x/y}\approx E_{QM}^{var}$ (see
Fig. 4).

\begin{equation}
\includegraphics[width=5in]{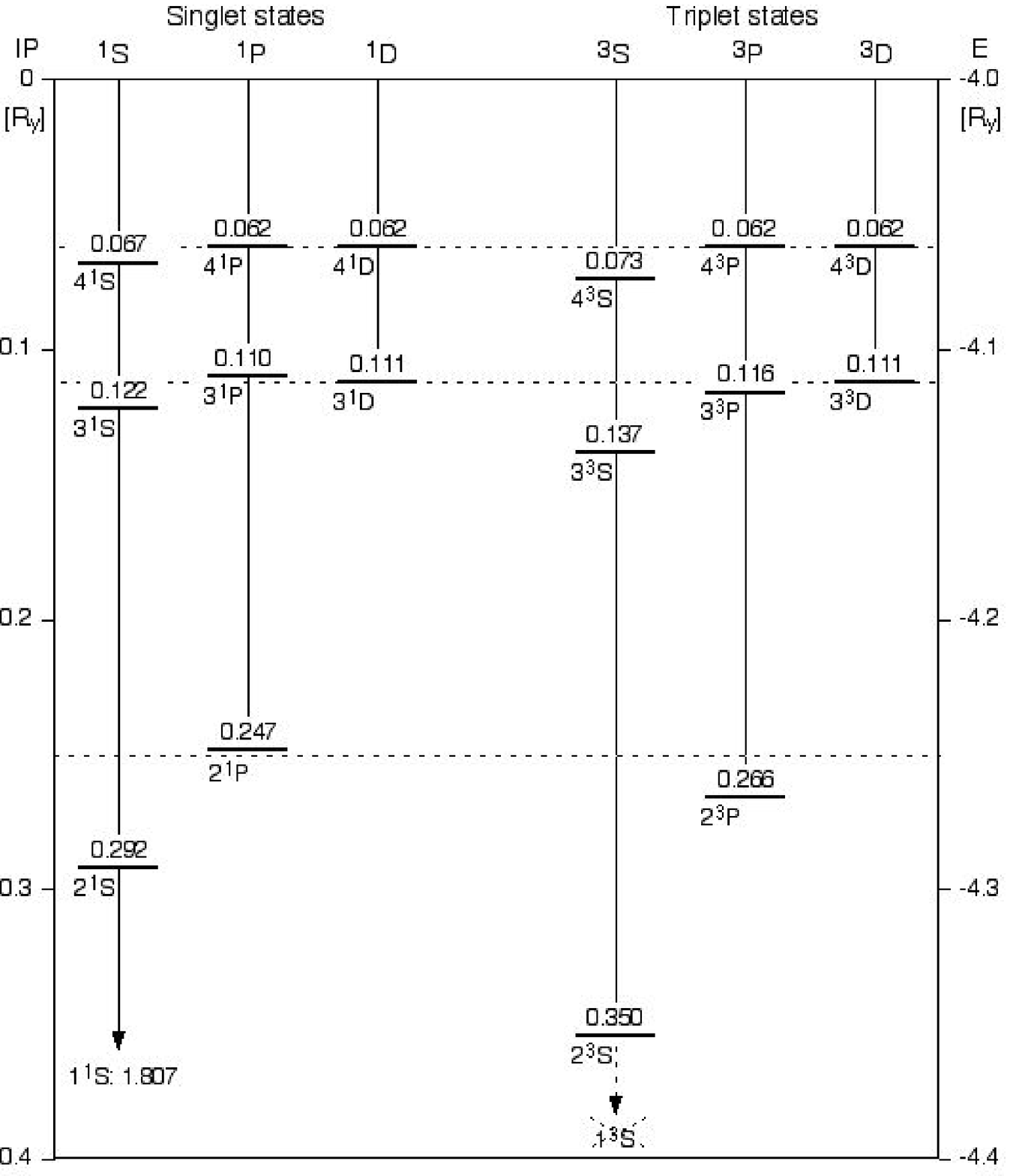}   \tag{}
\end{equation}
\begin{quote}
Fig. 5.  Experimental ionization potentials $IP$ (left scale) or energy levels $E$ (right scale) 
of helium in $R_{y}$ units.  The dotted horizontal lines represent the energy levels of hydrogen. \bigskip
\end{quote}

\section{EXCITED STATES}

All excited states of the $He$ atom consist of one electron in the $1s$
ground state and the other electron in an $nl$ excited state. [The lowest 
\textit{doubly} excited state $(2s,2s)$ has already a higher energy---above $%
2\times (-1R_{y})$---than the one-electron ion $He^{+}$ $(-4R_{y})$ and is
therefore unstable.] Figure 5 shows the double scheme of the $He$ energy
levels in terms of ionization potentials \textit{(}$IP$) with spectroscopic
notation $S,P,D$ for angular quantum numbers $l=0,1,2$. The energy level $%
E_{nl}$ of the $He$ atom (marked on the right scale of Fig. 5) is related to
the ionization potential (left scale) as 
\begin{equation}
E_{nl}=-4R_{y}-IP_{nl}\text{,}
\end{equation}

\noindent due to its atomic ground-state and excited-state contributions.

At first glance the level scheme of excited $He$ states in Fig. 5 may look
complicated. However, one quickly finds that many terms fall at the (dotted)
hydrogen levels. In these cases the excited electron experiences an
effective nuclear charge that results from the double charge of the $He$
nucleus and \textit{maximal\ shielding} by the $1s$ electron, 
\begin{equation}
Z_{H}=+2+(-1)=+1,
\end{equation}
that is, effectively a hydrogen nucleus.

The exceptions fall in three groups: the triplet $^{3}$S states, the singlet 
$^{1}$S states, and the triplet $^{3}$P states (see Fig. 5). Common to all
exceptions is that those excited terms fall \textit{beneath} the associated
hydrogen levels, $E_{n}(H)=-1R_{y}/n^{2}$. The reason is \textit{partial}
shielding, caused by penetration of the excited electron into the range of
the $1s$ electron orbit. The semiclassical orbits in Fig. 6 provide a visual
assessment, and qualitative explanation of the amount of partial shielding
for the three groups of exceptions. As in the case of the $He$ ground state,
we maintain the association of parallel and crossed Coulomb oscillations
with the excited triplet and, respectively, singlet states. If the excited
orbit is an $nl$ Sommerfeld ellipse, then the same association is extended
to the orientation of its \textit{major axis} relative to the $1s$ Coulomb
oscillator.

\begin{equation}
\includegraphics[width=3.4in]{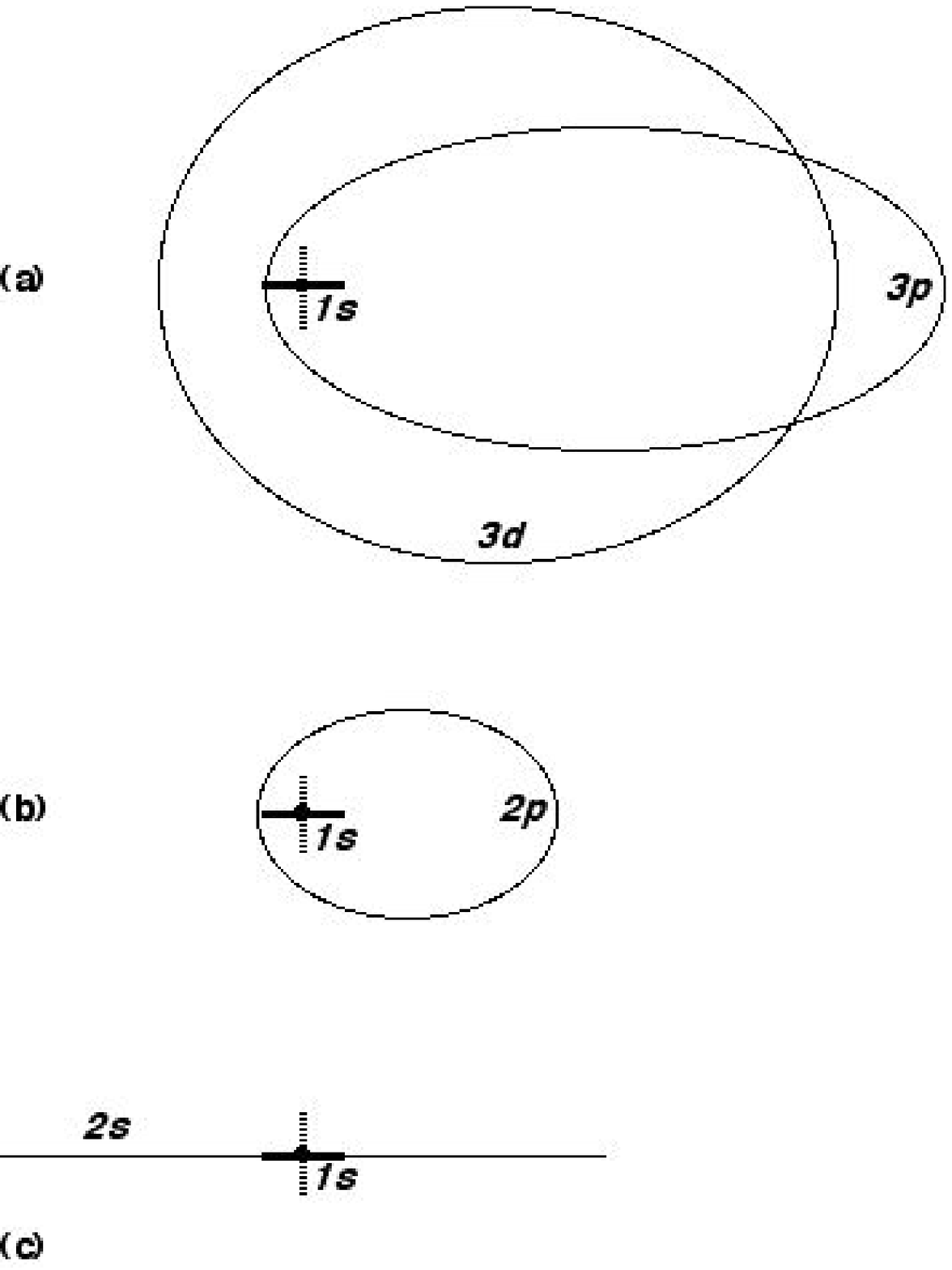}   \tag{}
\end{equation}
\begin{quote}
Fig. 6.  One-electron Coulomb oscillators and Sommerfeld ellipses about a $He$ nucleus, drawn to scale and arranged for easy comparison with Fig. 5.  The $1$s orbits are of a helium ion, $He^{+}$.  All other orbits are of a hydrogen atom, $H$, as if the nucleus was maximally shielded by $1s$. \bigskip
\end{quote}

An example of the first group of exceptions is the $2^{3}$S triplet state,
illustrated in Fig. 6(c) by \textit{parallel} Coulomb oscillators. The short
and fast Coulomb oscillation of the $1s$ ground-state electron is indicated
by the bold line; the ($\thicksim 8\times $ as) long and ($\thicksim 1/16$
as) slow oscillation of the (shielded) $2s$ electron by the thin line.
Outside the range of the $1s$ electron, the $2s$ electron experiences an
electric potential like from a hydrogen nucleus, $Z_{H}$. Inside the $1s$
range, it experiences only \textit{partial} shielding which results in
stronger binding (lower energy level). The same mechanism of $1s$-range
penetration and correspondingly lower energy levels also holds for the
parallel Coulomb oscillators with \textit{higher} quantum number $n$ (not
shown in Fig. 6), although to a lesser degree: The $2n^{2}$-times longer
amplitudes of the \textit{nth} (screened) Coulomb oscillator, compared to $%
1s $, causes a \textit{relatively} lesser lack of screening and leads to $%
n^{3}$S levels of lesser depth beneath the \textit{nth} hydrogen levels (see 
$3^{3} $S and $4^{3}$S in Fig. 5).

An example of the second group---the singlet $^{1}$S states---is $2^{1}$S,
illustrated in Fig. 6(c) by \textit{crossed} Coulomb oscillators. When now
the $2s$ electron oscillates \textit{perpendicularly} to the $1s$ electron,
it is less inside the (dashed) $1s$ range. Accordingly, the perpendicularly
oscillating $1s$ electron causes a \textit{lesser lack} of shielding as in
the previous case of parallel $1s$ oscillations, $E(n^{3}$S$)$ $<E(n^{1}$S$)$
$<E_{n}(H)$.

Examples of the third group---the triplet $^{3}$P states---are $2^{3}$P and $%
3^{3}$P, illustrated in Fig. 6(b,c), respectively. Passage of the $2p$ and $%
3p$ Sommerfeld orbits through the (bold) parallel $1s$ range is marginal,
causing only a small of lack of screening and level-lowering.

No passage of the $2p$ and $3p$ Sommerfeld orbits occurs through the \textit{%
crossed }(dashed) $1s$ range in Fig. 6(b,c). As a consequence of maximal
screening, the energies of the associated $2^{1}$P and $3^{1}$P singlet
states then fall (almost) at the hydrogen levels in Fig. 5. Finally, the $3d$
orbit in Fig. 6(c) passes far from the parallel and crossed $1s$ range. This
results in the $3^{3}$D and, respectively, $3^{1}$D terms at the $E_{3}(H)$
level.

\section{CONCLUSION}

The concept of the Coulomb oscillator overcomes impasses of the old quantum
theory not only with one-electron systems---orbital angular-momentum
hierarchy in the $H$ atom, stability of the $H_{2}^{+}$ molecule ion---but
also with the $He$ atom. Treating the motion of its \textit{two} electrons
as synchronous permits a reduction to a one-electron problem. Basic
distinctions arise for parallel and crossed Coulomb oscillations. They are
associated with the symmetry distinctions of Fermi-Dirac quantum statistics.
With regard to the average electron-electron distance, combined with
screening effects, the parallel and crossed Coulomb oscillations emulate
quantum-mechanical symmetry constraints and accordingly yield similar
ground-state and excited-state energy levels of the $He$ double spectrum.

\bigskip \noindent

{\Large \noindent }\noindent \noindent \noindent {\Large ACKNOWLEDGMENTS}

\noindent I thank Ernst Mohler for valuable discussions and for bringing the
cross-delayed oscillator to my attention. Thanks also to Duane Siemens and
Preston Jones for discussions and help with computers.\bigskip \bigskip
\pagebreak

\noindent Table I. Ground-state energy of $He$ and two-electron ions:
Experimental values $E_{\exp t}$ (Bethe and Salpeter 1957) and calculations
with contralinear ($-xx$), cross-synchronous ($xy$), and cross-delayed ($x/y$%
) Coulomb oscillators, and one-parameter $QM$ variation. Relative deviations
are denoted by $\Delta .$\smallskip

\noindent \noindent 
\begin{tabular}{ccccccccc}
Ion & Z & $\frac{E_{\text{\textit{expt}}}}{[R_{y}]}$ & $\frac{E_{-xx}}{
[R_{y}]}$ & $\frac{\Delta _{-xx}}{[\%]}$ & $\frac{E_{xy}}{[R_{y}]}$ & $\frac{
E_{x/y}}{[R_{y}]}$ & $\frac{\Delta _{x/y}}{[\%]}$ & $\frac{E_{QM}^{var}}{
[R_{y}]}$ \\ 
$H^{-}$ & 1 & -1.06 & -1.13 & 7.6 & -0.84 & -0.99 & 5.7 & -0.95 \\ 
$He$ & 2 & -5.81 & -6.13 & 5.5 & -5.42 & -5.77 & 0.6 & -5.70 \\ 
$Li^{+}$ & 3 & -14.56 & -15.13 & 3.9 & -14.01 & -14.57 & 0.0 & -14.45 \\ 
$Be^{2+}$ & 4 & -27.31 & -28.13 & 3.0 & -26.59 & -27.36 & 0.2 & -27.20 \\ 
$B^{3+}$ & 5 & -44.06 & -45.13 & 2.4 & -43.18 & -44.16 & 0.2 & -43.95 \\ 
$C^{4+}$ & 6 & -64.82 & -66.13 & 2.0 & -63.76 & -64.94 & 0.2 & -64.70 \\ 
$N^{5+}$ & 7 & -89.58 & -91.13 & 1.8 & -88.35 & -89.74 & 0.2 & -89.45 \\ 
$O^{6+}$ & 8 & -118.34 & -120.13 & 1.5 & -116.94 & -118.54 & 0.2 & -118.20
\end{tabular}

\end{document}